\documentclass[]{IEEEphot}

\jvol{xx}
\jnum{xx}
\jmonth{June}
\pubyear{2009}

\usepackage{epsf}
\usepackage{graphicx}

\usepackage{amsmath}
\usepackage{algorithm}
\usepackage[noend]{algpseudocode}
\usepackage{array}
\usepackage{multirow}
\usepackage{booktabs}

\usepackage[noadjust]{cite}

\begin{document}

\title{Seamless Rate Adaptation at Receiver Side for Indoor Visible Light Communications by Using Raptor Codes}

\author{Cenk~Albayrak$^{\textbf{1}}$,~Kadir~Turk$^{\textbf{1}}$,~Emin~Tugcu$^{\textbf{2}}$,~Cemaleddin~Simsek$^{\textbf{2}}$~and~Ayhan~Yazgan$^{\textbf{1}}$}

\affil{\normalfont $^{\textbf{1}}$Department of Electrical and Electronics Engineering, Karadeniz Technical University,\\ Trabzon 61080, Turkey\\
	$^{\textbf{2}}$Department of Electronics and Communication Engineering, Karadeniz Technical University,\\ Trabzon 61830, Turkey} 


\maketitle


\begin{receivedinfo}%
This work was supported by the Scientific and Technological Research Council of Turkey (TUBITAK) Project Number 215E308. This paper was presented in part at 40th International Conference on Telecommunications and Signal Processing (TSP 2017).
\end{receivedinfo}


\begin{abstract}
Visible light communication (VLC), which uses LEDs for both illumination and communication purposes, has recently emerged as either an alternative or a support system for radio frequency (RF) communication.
VLC channel inside a room is highly variable since it is pretty much depends on optical line of sight (LOS) and field of view (FOV) and it causes the received signal to vary dramatically. Therefore, to achieve spectral efficient communication, using rate adaptation methods are essential.
In this study, we propose a rateless codes based rate adaptation method which adjusts transmission rate at receiver side and does not require channel state information (CSI) at the transmitter side. 
Spectral efficiency of proposed method is analyzed and compared with adaptive modulation and coding (AMC) rate adaptation method widely used in literature. For simulation work, we consider a realistic communication system model by using commercially available LEDs and also fulfilling illumination requirements. 
To overcome multi-path effects, optical orthogonal frequency-division multiplexing (OFDM) is used.
Simulation results indicate that proposed rate adaptive method can achieve seamless rate adaptation and has higher spectral efficiency compared to AMC in presence of significant signal quality variations caused by mobility in a model room.

\end{abstract}


\section{Introduction}
\label{Introduction}

In recent years, usage of light emitting diodes (LEDs) for illumination purposes have increased rapidly because of their higher lumen/watt values compared to traditional lighting devices. Also, they provide environment friendly illumination with their long life spans. Beside these benefits, LEDs have high speed switching features which makes them a good candidate for short range, indoor wireless optical communication front end. This technology is known as visible light communication (VLC) which uses LEDs for both illumination and communication purposes~\cite{IEEE_Standard, VLC_mobile,VLC_uysal,VLC_Noma,VLC_survey,VLC_analog}. Compared to radio frequency (RF) communication, VLC systems provide higher security without generating RF interference and have a unlicensed bandwidth. Recently, VLC has emerged as either an alternative or a support system for some of the scenarios carried by RF communication systems. With a widespread utilization of LEDs, VLC can be quickly adapted to wireless network at hotspots such as supermarkets, shopping centre, aircraft cabins and etc.

From communication perspective, VLC system contains LEDs and photo-diodes (PD) as transmitter and receiver front ends, respectively. Communication channel between LEDs and PDs is highly variable and pretty much depends on optical line of sight (LOS) and field of view (FOV). The received signal strength varies dramatically inside a room where VLC is implemented \cite{VLC_mobile,VLC_analog}.
In most works related to VLC, it is assumed that the transceiver units are located at fixed points. 
However, mobility is one of the most desired features by users. 
To achieve high spectral efficiency for the VLC systems where the channel conditions vary significantly, using rate adaptation methods are essential~\cite{CenkKadir_TSP,VLC_LEDBased,LLR-ADM}. 

To improve spectral efficiency, adaptive modulation and coding (AMC) method is widely used for RF communication systems. Systems using AMC method dynamically adjust code rate and modulation depth at transmitter side according to channel state information (CSI) sent by receiver \cite{LLR-ADM,LLR-ADM_glo}. 
However, receiver may not provide an accurate CSI or CSI might be delayed which significantly impairs the performance of system using AMC. 
In particular, if the channel condition changes fast, such a feedback will require high overhead for channel estimation training and also cost much feedback channel resource. Furthermore, if the channel varies faster, the feedback of CSI may not useful to the transmitter in the AMC systems.
Also, AMC can only achieve staircase throughput rates due to limited number of fixed code rates and modulation depth pairs combinations. Additionally, AMC systems are not applicable for broadcast/multicast applications with a single transmitter and multiple receivers because CSIs of different receivers are likely to be different.

Another method to improve spectral efficiency employed in RF systems is using rateless codes based rate adaptation \cite{LLR-ADM,LLR-ADM_glo,Rateless_broad}. Rateless codes such as Luby transform (LT) \cite{LTCodes} or Raptor codes \cite{RaptorCodes} do not limit the system with limited number of fixed code rates as in AMC, on the contrary they provide a limitless stream of encoded symbols which are independent and identically distributed (i.i.d.), on the fly or in advance with continuous number code rate \cite{FixedRate}. In this method, encoded symbols are transmitted until the receiver sends an acknowledgment to the transmitter after successful decoding. As a result, transmission code rate can be adjusted according to channel conditions at the receiver side instead of transmitter, i.e. rateless codes do not require CSI feedback. Therefore, rate adaptation with rateless codes can minimize the feedback requirement to one-bit acknowledgment and achieve seamless rate adaptation which provides higher spectral efficiency compared to AMC method.

In \cite{VLC_rate_adap,VLC_rate_adap2,VLC_rate_adap3,VLC_rate_adap4} adaptive modulation (AM) method is used to improve spectral efficiency of VLC system, however AM method also requires CSI feedback from receiver and has staircase rates as in AMC. In \cite{VLC_analog}, a hybrid VLC-RF system with analog rateless codes is proposed as a rate adaptation framework. However, in \cite{VLC_analog} illumination is not considered and the system is evaluated over approximated VLC channel model. Also, encoding and decoding complexities of analog rateless codes are much higher than conventional rateless codes, which puts these codes back using in practical systems. 

In this study, we propose a rate adaptive system with Raptor codes for VLC system to improve spectral efficiency where channel conditions vary significantly. To overcome inter-symbol interference (ISI) caused by multi-path effects, we use optical orthogonal frequency-division multiplexing (OFDM) which has been widely used for VLC systems \cite{VLC_analog,DCO_ACO_OFDM2}. On contrast to literature using AM or analog rateless codes methods, we consider a realistic communication system model by using commercially available LEDs (Cree Xlamp XB-H) and also fulfilling illumination requirements in our analysis. Considered simulation system model contains LED armatures which simultaneously transmits same data where receiver is mobile inside a room. Simulation results indicate that proposed rate adaptive method can achieve seamless rate adaptation in presence of significant signal quality variations via mobility. 

The remainder of this paper is organized as follows. In Section~\ref{Visible Light Communications}, VLC channel structure and optical OFDM schemes is introduced. Calculation of signal-to-noise ratio (SNR) for VLC is \text{explained} in same section. Rateless codes and its soft decoding process are presented in Section~\ref{Rateless Codes}. In Section~\ref{Proposed Rate Adaptation Scheme}, we elaborate proposed rate adaptive VLC scheme. Simulation results that contain comparison of proposed method with AMC scheme are given in Section~\ref{Simulation Results}. Finally some conclusion is drawn in Section~\ref{Conclusion}.


\section{Visible Light Communications}
\label{Visible Light Communications}
In this section, the essential components of the proposed VLC scheme are briefly explained, starting with VLC channel structure, relation between optical SNR vs electrical SNR and optic OFDM techniques for VLC channel.

\subsection{VLC Channel Structure}
In this part of the study, a ray tracing based channel model which is introduced in \cite{VLC_indoor} and has been widely recognized in literature for VLC systems is utilized \cite{VLC_survey}. To obtain total channel impulse response (CIR) for this channel model, combinations of individual CIRs between $n$-th source and receiver by direct and reflected propagations are calculated \cite{VLC_indoor}. For $k$ reflections, the impulse response can be obtained as follows,
\begin{align}\label{h_CIR}
	h(t)=\sum_{n=1}^{N_{t}}\sum_{i=0}^{k}h^{(i)}(t;\Phi_{n})
\end{align}
where $N_{t}$ is the total number of LED sources and $\Phi_{n}$ is the power spectral distribution of $n$-th source, which describes the radiant power per unit wavelength of an illuminator. The individual response of $n$-th source experiencing the $i$-th reflection is calculated as follows,
\begin{align}\label{hk_CIR}
	h^{(i)}(t;\Phi_{n})=\int_{S} \bigg[L_{1}L_{2} \ldots L_{i+1} \ \Gamma_{n}^{(i)} \ rect \bigg( \frac{\theta_{i+1}}{FOV} \bigg) \times \delta \bigg( t-\frac{d_{1}+d_{2}+ \ldots +d_{i+1}}{c} \bigg) \bigg] dA_{ref}, \ i \geq 1
\end{align}
where $S$ is the area of all reflection surfaces, $L$ is the path-loss term for each path, $\Gamma_{n}^{(k)}$ stands for the $i$-th reflection power of $n$-th source, $rect(.)$ represents the rectangular function, $\theta$ is the incidence angle and the field of view (FOV) is the acceptance angle of receiver. $t$, $d$, $c$ and $A_{ref}$ represent time, the distance between light source and receiver, the speed of light and the effective reflection area, respectively. Calculations of $L$, $\Gamma$ and $rect(.)$ are examined with details in \cite{CenkKadir_TSP} and \cite{VLC_indoor}. In line of sight, i.e. when $i=0$, the individual response of $n$-th source is calculated with Equ.~(\ref{h0_CIR}).
\begin{align}\label{h0_CIR}
	h^{(0)}(t;\Phi_{n})=L_{0} \ \Gamma_{n}^{(0)} \ rect \bigg( \frac{\theta_{0}}{FOV}\bigg) \times \delta \bigg( t-\frac{d_{0}}{c} \bigg)
\end{align} 


\subsection{Electrical Signal to Noise Ratio}
\begin{figure}[!ht]
	\centering
	\includegraphics[width=35pc]{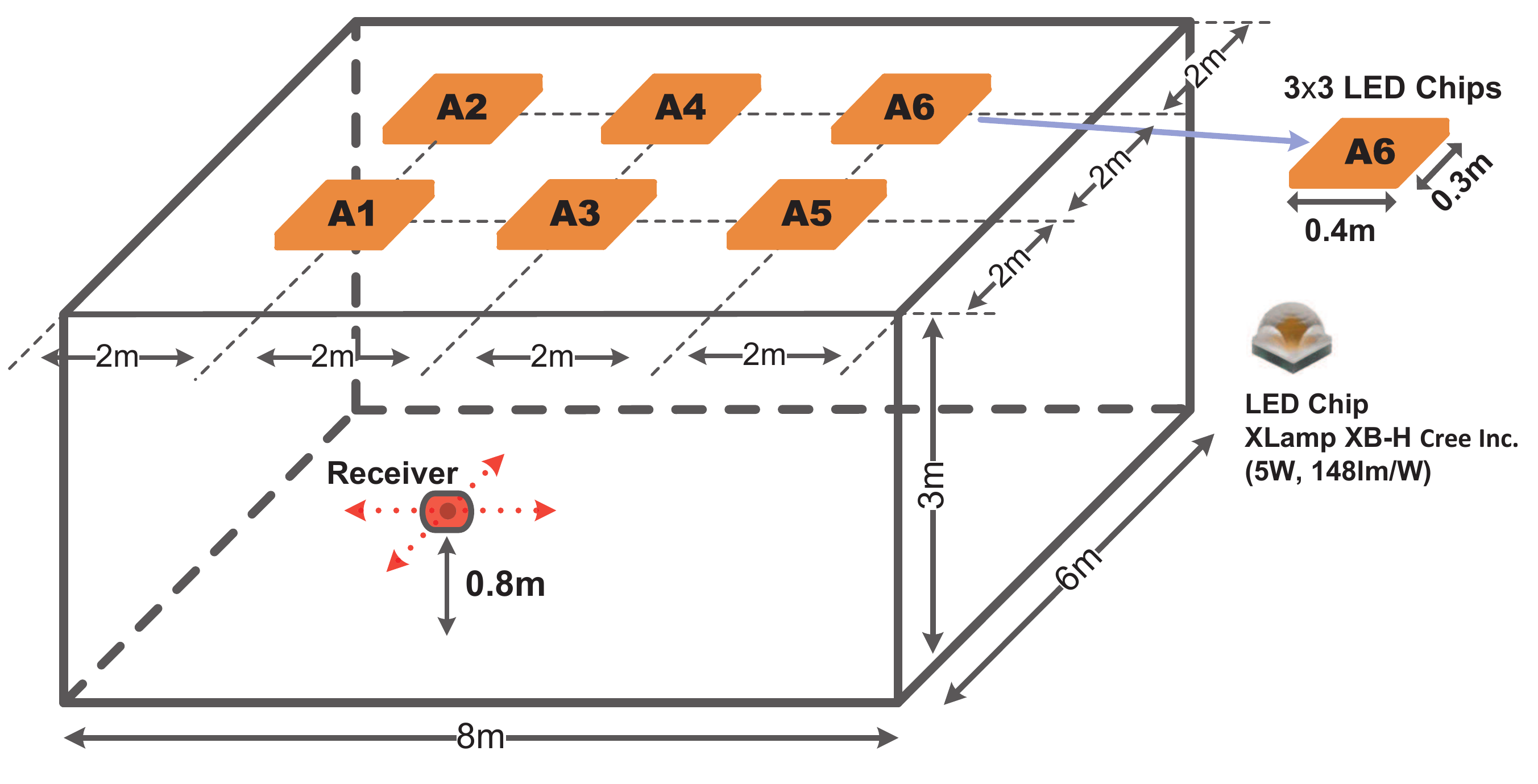}
	\caption{VLC model room size of 8m x 6m x 3m. Commercially available LEDs (Cree Xlamp XB-H) are used for both illumination and communications purpose. An LED luminary consists of 9 LED chips.}
	\label{fig_room}
\end{figure}

VLC channel can be modeled as a baseband linear time-invariant system with input optical power (intensity) $x(t)$, channel impulse response $h(t)$ as follows \cite{VLC_Infrared}, 
\begin{align}\label{VLC_Channel}
	y(t)=Rx(t)\otimes h(t)+n(t)
\end{align}
where $y(t)$ represents the received signal, $R$ is the photo-diode (PD) responsivity, $n(t)$ is additive white Gaussian noise (AWGN) and the symbol $\otimes$ stands for convolution. In general, a typical VLC communication system includes a pair of electro-optic transceivers at both transmitter and receiver. Transmitter which consists of several electronic circuits such as driver, amplifier and modulator, converts the electrical signal into optical domain ($x(t)$) using LEDs. A photo-diode based electronic circuit consists of the appropriate amplifier and demodulator is located at the receiver side where the optical signal is converted to electrical signal ($y(t)$) properly. In this study, we assume a FET based transimpedance preamplifier is used at the receiver side. Since the channel input represents the optical power, $x(t)$ must be real and positive. Average transmitted optical power $P_{t}$ is given by $P_{t}=\lim_{T \rightarrow \infty} \frac{1}{2T} \int_{-T}^{T} x(t)dt$ rather than usual time-average of $|x(t)|^{2}$, which is appropriate when $x(t)$ represents amplitude \cite{VLC_Adap}. Let $P_{r}$ represents average received optical power, electrical SNR is given as follows \cite{VLC_Funda},
\begin{align}\label{d_pr}
	SNR=\frac{(RP_{r})^{2}}{\sigma_{n}^{2}}
\end{align}
where noise variance consists of shot and thermal noise variances. It is given by $\sigma_{n}^{2}=\sigma_{shot}^{2}+\sigma_{thermal}^{2}$ and the noise variances are obtained by following equations,
\begin{align}\label{d_shot_thermal}
	&\sigma_{thermal}^{2}=\frac{8\pi \kappa T_{k}}{G}C_{p}AI_{2}B^{2}+\frac{16\pi^{2} \kappa T_{k} \Gamma_{F}}{g_{m}}C_{p}^{2}A^{2}I_{3}B^{3},& \nonumber \\
	&\sigma_{shot}^{2}=2qRP_{r}B+2qI_{bg}I_{2}B&
\end{align}
where $\kappa$ denotes the Boltzmann's constant, $T_{k}$ is absolute temperature, $G$ is open-loop voltage gain, $C_{p}$ is fixed capacitance per unit area, $A$ is area of PD, $I_{2}$ and $I_{3}$ represent noise-bandwidth factors, $B$ is equivalent noise bandwidth, $\Gamma_{F}$ is FET channel noise factor and $g_{m}$ is FET transconductance, $q$ is electronic charge and $I_{bg}$ is background current \cite{VLC_Adap,VLC_Funda}. Average received optical power is obtained by $P_{r}=H(0) \ P_{t}$, where $H(0)$ represents channel DC gain and is given with Equ.~(\ref{DC_gain}) \cite{VLC_Infrared}.
\begin{align}\label{DC_gain}
	H(0)=\int_{-\infty}^{\infty} h(t) dt
\end{align}
\indent As mentioned in Section~\ref{Introduction}, to compare achieved throughputs and spectral efficiecies of AMC and proposed rate adaptive system, an empty room with 8m x 6m x 3m size is modeled as shown in Fig.~\ref{fig_room}. Commercially available six LED armatures (Cree Xlamp XB-H) are located on ceiling at locations of A1 (2.0m, 2.0m, 3.0m), A2 (2.0m, 4.0m, 3.0m), A3 (4.0m, 2.0m, 3.0m), A4 (4.0m, 4.0m, 3.0m), A5 (6.0m, 2.0m, 3.0m) and A6 (6.0m, 4.0m, 3.0m). One LED armature consists of 9 LED chips and each chip radiates 5 W with a viewing angle of 110$^{o}$. This design yields an average illumination level of 362 lux which satisfies the typical illumination levels for a laboratory environment \cite{AydinlatmaStandart}.

\begin{figure}[!ht]
	\centering
	\includegraphics[width=35pc]{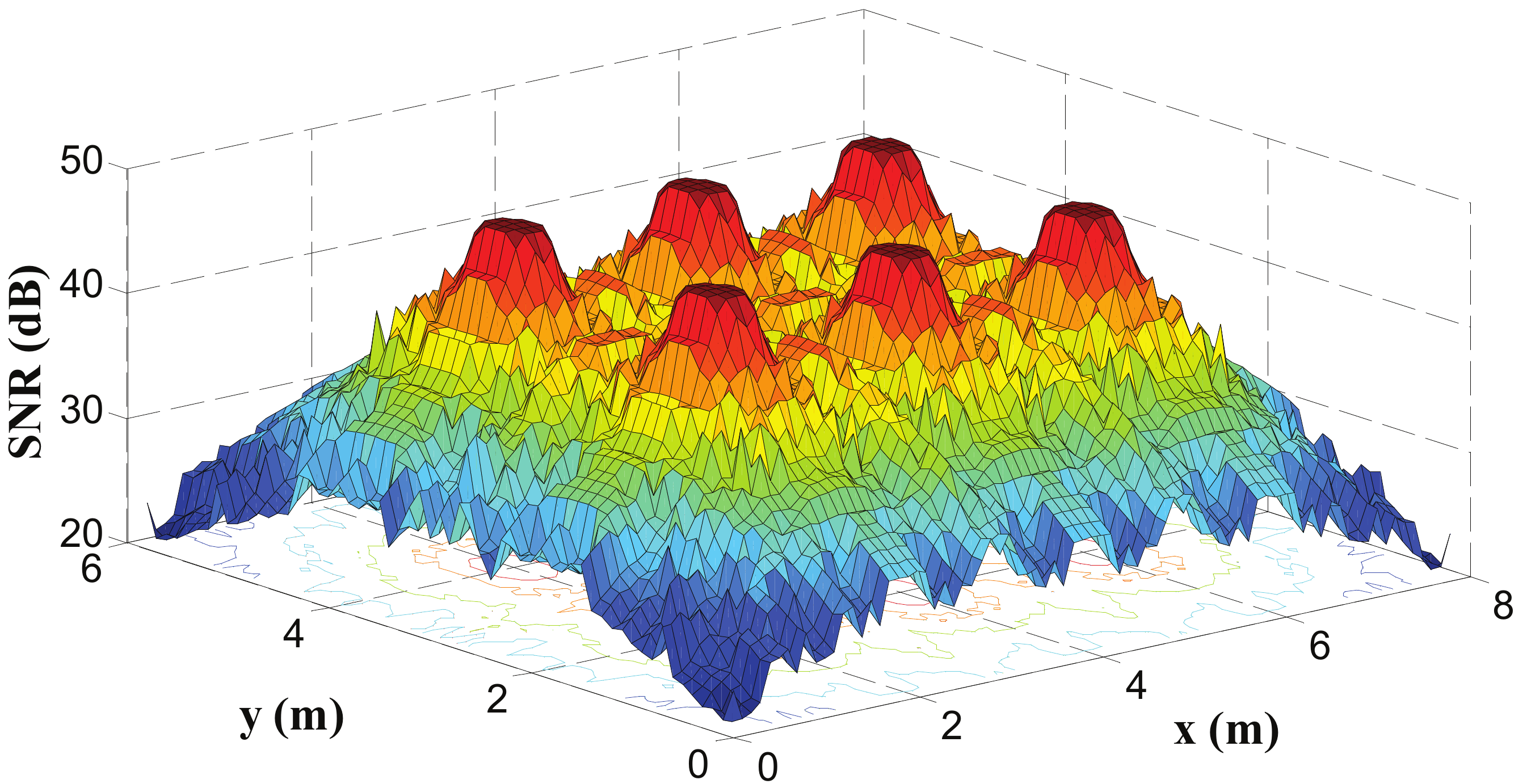}
	\caption{Distribution of SNR including direct and reflected light in the considered model room}
	\label{fig_snr_distribution}
\end{figure}

To obtain SNR distribution for the model room, a laboratory set up is modeled by using the channel parameters as given in Table~\ref{tab_noise} \cite{VLC_Funda} where the detector is mobile at 0.8m height and a fixed 90$^{o}$ angle. According to these conditions, obtained SNR distribution is given in Fig.~\ref{fig_snr_distribution}. The results show that, when the receiver is mobile, observed SNR variation is between 20.42 dB and 47.61 dB at detector. Because of this significant variation of SNR, a rate adaptive system must be used for spectral efficient transmission.
\begin{table}[!ht]
	\centering
	\caption{Parameter Values for the shot and thermal noise variances \cite{VLC_Funda}}
	\label{tab_noise}
	\begin{tabular}{@{}lll@{}}
		\toprule
		Symbols      & Descriptions             & Values                \\ \midrule
		R            & PD responsivity          & $0.54$ {[}A/W{]}      \\
		$A$          & Area of PD               & $1.0$ {[}cm$^2${]}    \\
		$\kappa$     & Boltzmann's constant     & $1.38$                \\
		$T_{k}$      & Absolute temperature     & $298$ {[}K{]}         \\
		$G$          & Open-loop voltage gain   & $10$                  \\
		$I_{2}$      & Noise-bandwidth factor   & $0.562$               \\
		$I_{3}$      & Noise-bandwidth factor   & $0.0868$              \\
		$I_{bg}$     & Background current       & $5100$ {[}$\mu$A{]}   \\
		$C_{p}$      & Fixed capacitance        & $112$ {[}pF/cm$^2${]} \\
		$\Gamma_{F}$ & FET channel noise factor & $1.5$                 \\
		$g_{m}$      & FET transconductance     & $30$                  \\
		$q$          & Electronic charge        & $1.602$               \\
		$B$          & Bandwidth                & $100 $  {[}MHz{]}        \\ \bottomrule
	\end{tabular}
\end{table}

\subsection{Optical OFDM}
\label{Optical OFDM}
OFDM is extensively used in broadband communication systems as it is an effective solution to combat ISI caused by multi-path effects of the communication channel \cite{VLC_analog}. Recently, there has been a growing interest in employing OFDM for VLC systems to overcome dispersion effects of the VLC channel \cite{VLC_analog,DCO_ACO_OFDM2}. In intensity modulation/direct detection (IM/DD) based VLC systems, only non-negative real valued signals can be modulated through light intensity, because the intensity of light can not be negative. Since OFDM signals are bipolar and complex, conventional OFDM techniques can not be directly applied to VLC systems. Therefore, several forms of OFDM such as DC biased optical OFDM (DCO-OFDM) and asymmetrically clipped optical OFDM (ACO-OFDM) are proposed for VLC systems \cite{DCO_ACO_OFDM1}. In OFDM IM/DD systems, a real OFDM signal can be generated using the Hermitian symmetry property of inverse Fourier transform. The transmitted signal can be made positive either by adding a DC bias, as in DCO-OFDM, or by clipping at zero as in ACO-OFDM \cite{DCO_ACO_OFDM1}. Because of DC biasing, DCO-OFDM is less efficient than ACO-OFDM in terms of average optical power, while ACO-OFDM is less efficient than DCO-OFDM in terms of bandwidth because of fewer subcarrier usage.

In \cite{DCO_ACO_OFDM2}, it is shown that for relatively small constellations such as 4-QAM, 16-QAM, 64-QAM and 256-QAM, ACO-OFDM is more efficient than DCO-OFDM in terms of average optical power. However, for larger constellations such as 1024-QAM and 4096-QAM, the effect of bandwidth efficiency in ACO-OFDM is dominant and DCO-OFDM gives better performance than ACO-OFDM. 
Obtained results given in Section~\ref{Visible Light Communications} show that SNR level is high at the receiver for the considered model room and thus a high-order modulation type should be selected in order to obtain high spectral efficiency. Furthermore, DCO-OFDM technique has a more simpler structure than ACO-OFDM \cite{DCO_ACO_OFDM2}. Therefore, DCO-OFDM technique has been used in this study.\\
\indent In DCO-OFDM, the DC bias level denoted by $B_{DC}$ is set relative to the standard deviation of $s(t)$ which is the OFDM signal drives the LEDs given as follows \cite{DCO_ACO_OFDM1},
\begin{align}\label{DC_bias}
	B_{DC}=\mu \sqrt{E\{s(t)^{2}\}}
\end{align}
where $\mu$ is a proportionality constant and $B_{DC}$ is obtained as a bias of $10$log$_{10}(\mu^{2}+1)$dB. After the addition of $B_{DC}$ to the OFDM signal, any remaining negative peak is clipped at zero and obtained signal is transmitted by LEDs \cite{DCO_ACO_OFDM2}.

The effect of DC bias in DCO-OFDM is investigated in different studies \cite{VLC_rate_adap4}, \cite{DCO_ACO_OFDM1}. Results given in these studies show that a lower DC bias achieves a higher spectral efficiency if clipping noise effect doesn't dominate. In other words, a lower DC bias can cause to clipping noise. Therefore, optimum DC bias that the clipping noise can be ignored should be used to achieve a higher spectral efficiency in DCO-OFDM. 


\section{Rateless Codes}
\label{Rateless Codes}
LT and Raptor codes are members of rateless codes family \cite{FixedRate}. LT codes are the first practical realizations of rateless codes \cite{LTCodes}. An LT encoder generates a limitless stream of encoded symbols, which are independent and identically distributed (i.i.d.), on the fly or in advance \cite{LLR-ADM}. On the other hand, Raptor codes are an extended version of LT codes with linear encoding and decoding complexities \cite{RaptorCodes}. A Raptor codes consist of an outer fixed rate LDPC precode and an inner weakened LT code which has less computational cost \cite{RaptorLowSNR}. At decoder, source data can be reconstructed from arbitrarily collected encoded symbols.\\
\indent Message-passing based belief propagation (BP) decoding algorithm is used for soft decoding of rateless codes over the noisy channels \cite{FixedRate}. Tanner graph efficiently visualizes the decoding process of rateless codes and contains two types of nodes named as the check-node (CN) and variable-node (VN) \cite{Tanner}. The log-likelihood ratio (LLR) messages are passed along each edge in the tanner graph between neighboring VNs and CNs iteratively \cite{LLR-ADM}. Decoding for Raptor codes starts with LT decoding. Let $m_{o,i}$ and $m_{i,o}$ represent outgoing LLR messages from CN $o$ to VN $i$ and from VN $i$ to CN $o$, respectively. At LT decoding process, message update equations are formulated as follows, 
\begin{align}\label{mC_mV_LT}
	&m_{o,i}^{(l)} = 2tanh^{-1} \Bigg[tanh\left(\frac{m_{c}}{2} \right) \prod_{i' \neq i} tanh \left(\frac{m_{v_{i',o}}^{(l-1)}}{2} \right) \Bigg]& \\
	&m_{i,o}^{(l)}=\sum_{o' \neq o}m_{o',i}^{(l)}&
\end{align}
where $m_{c}$ stands for LLR values of the codewords from channel, superscript $l$, $tanh(\cdot)$ and $tanh^{-1}(\cdot)$ denotes the iteration number, hyperbolic tangent and inverse hyperbolic tangent operations, respectively \cite{LLR-ADM}. After running LT decoder for predetermined number of iterations, CN messages in every VN are combined and set as the messages from channel for corresponding VN of LDPC precode as $m_{i,v}=\sum_{o}m_{o,i}^{(l)}$. Let $m_{v,c}$ and $m_{c,v}$ denote outgoing LLR messages from CN $c$ to VN $v$ and from VN $v$ to CN $c$, respectively. At LDPC decoding process, message update equations are formulated as follows \cite{FixedRate}, 
\begin{align}\label{mV_mC_LDPC}
	&m_{v,c}^{(l)}=m_{i,v}+\sum_{c' \neq c}m_{c',v}^{(l-1)}& \\
	&m_{c,v}^{(l)} = 2tanh^{-1} \Bigg[\prod_{v' \neq v} tanh\left(\frac{m_{v',c}^{(l)}}{2} 	\right) \Bigg]&
\end{align}	
After running LDPC decoder for predetermined number of iterations, decision process is done as follows, 
\begin{align}\label{Mv_Decision}
	m_{v}= m_{i,v}+\sum_{c}m_{c,v}^{(l)}, \quad \quad \hat{m}_{v}= \left\{ 
	\begin{array}{ll}
		0 & m_{v} \geq 0\\
		1 & m_{v} < 0	
	\end{array} \right.
\end{align}
where $\hat{m}_{v}$ denotes estimated bits of decoder output.


\section{Proposed Rate Adaptation Scheme for VLC}
\label{Proposed Rate Adaptation Scheme}

\begin{figure}[!ht]
	\centering
	\includegraphics[width=36pc]{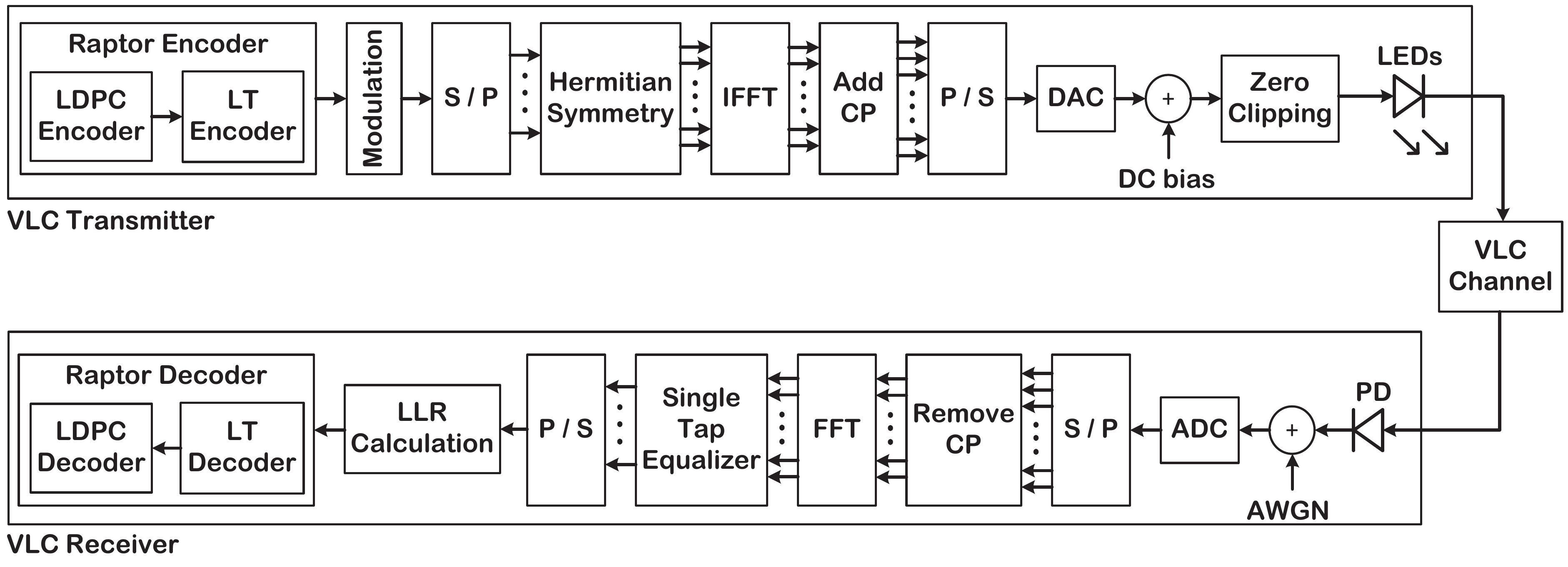}
	\caption{Block diagram of proposed rate adaptive system with Raptor codes for VLC}
	\label{fig_dco_ofdm}
\end{figure}

Proposed rate adaptive VLC scheme and system parameters are explained in this section. Block diagram of the proposed scheme is shown in Fig.~\ref{fig_dco_ofdm}. At the encoding process, information bits are first pre-coded by a rate of 0.98 right-Poisson and left-regular LDPC code of left degree four \cite{RaptorCodes} to produce intermediate bits and then encoded bits are produced from intermediate bits by LT encoder with following degree distribution as widely used in the literature \cite{RaptorCodes}.
\begin{align}\label{degreeDist}
	\Omega(x)= \ &0.004807x+0.496472x^{2}+0.166912x^{3}+0.073374x^{4}+0.082206x^{5}+0.057471x^{8}& \\
	&+0.035951x^{9}+0.001167x^{18}+0.054305x^{19}+0.018235x^{65}+0.0091x^{66}& \nonumber
\end{align}

At transmitter side, encoded bits are modulated at the highest modulation depth allowed by VLC channel in order to achieve the possible highest spectral efficiency. OFDM symbols are then constituted as detailed in Section~\ref{Optical OFDM} and an appropriate DC bias is added to OFDM signal. After the addition process, any remaining negative peak is clipped at zero and the resulting signal is transmitted by LEDs as shown in Fig.~\ref{fig_dco_ofdm}.

In this system, VLC transmitter keeps sending symbols modulated by chosen modulator and the entire decoding process is repeated by increasing number of encoded symbols at output of Raptor encoder until VLC reciever collects enough symbols for successful decoding. At the decoding process, LT code is firstly decoded and the decoding of LDPC precode is performed as detailed in Section~\ref{Rateless Codes}. After successful decoding, the receiver sends one-bit acknowlegment to VLC transmitter via feedback channel \cite{RaptorLowSNR}. Thus, proposed scheme can automatically adapt to channel conditions by adjusting code rate at receiver side instead of transmitter side as in AMC method. 


\section{Simulation Results}
\label{Simulation Results}
For simulation work, a model room as detailed in Section~\ref{Visible Light Communications} for a realistic VLC channel is considered. In the model room we used same materials as in \cite{VLC_indoor} (walls, floor and ceiling). With this scenario, throughput performances of proposed rate adaptive scheme is compared with AMC in presence of mobility. For all simulation works, we use M-QAM modulation as widely chosen for DCO-OFDM in the literature \cite{DCO_ACO_OFDM2,DCO_ACO_OFDM1} and assume that OFDM symbols consist of 1024 subcarriers and 16 cyclic prefix. Also, it is supposed that all LEDs simultaneously transmit same data.

\begin{figure}[!ht]
	\centering
	\includegraphics[width=35pc]{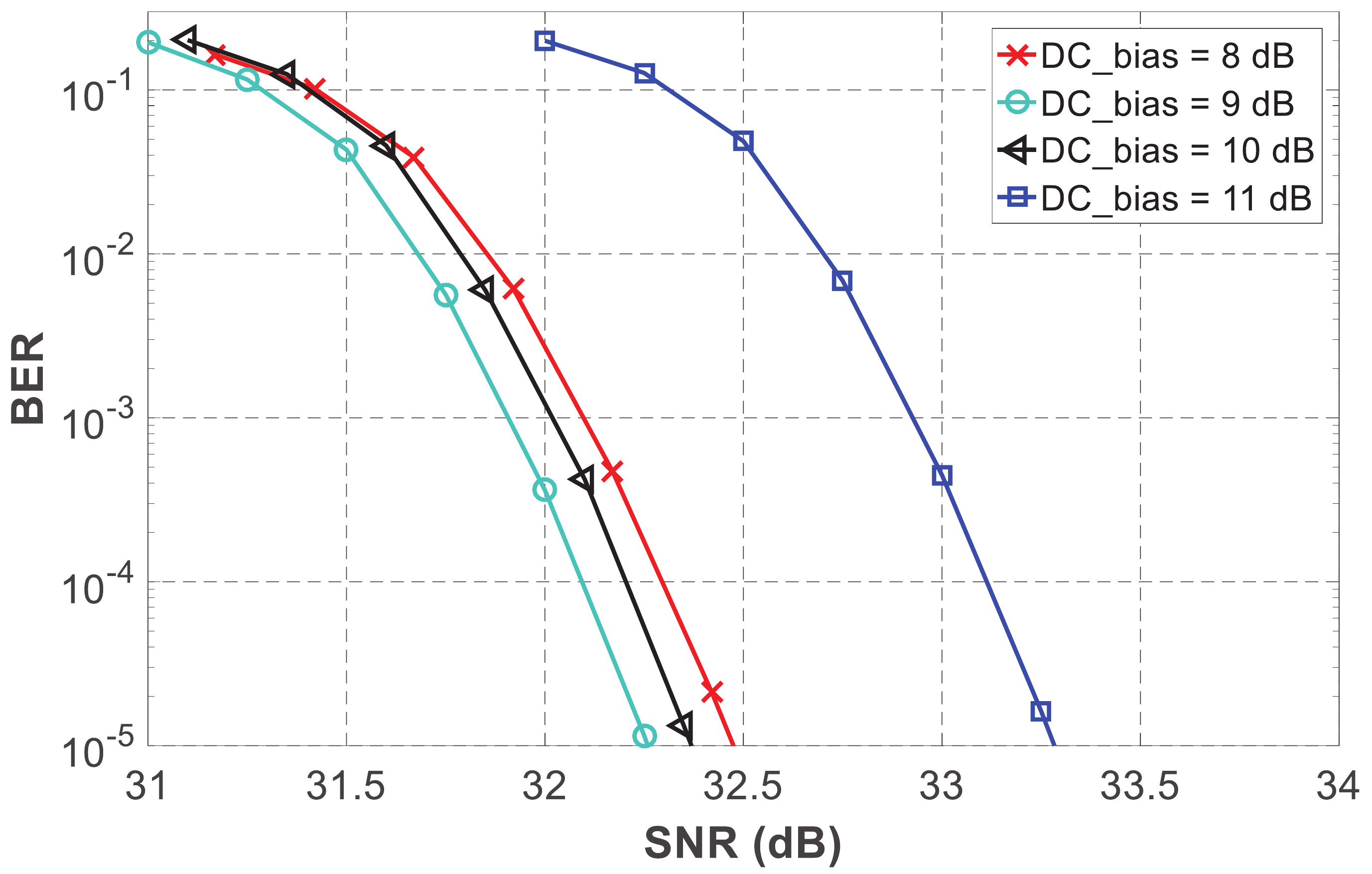}
	\caption{BER curves for different DC bias values when the receiver location is (2.6m,~0.7m,~0.8m) inside the model room and r is choosen 0.5.}
	\label{fig_snr_ber}
\end{figure}

As mentioned in Section 2, DC bias is an important parameter for DCO-OFDM which has a direct effect on performance of proposed method. To be able to choose optimum DC bias, BER performances with different DC bias levels are calculated and results are shown in Fig.~\ref{fig_snr_ber} for \text{4096-QAM} modulation depth. According to Fig.~\ref{fig_snr_ber}, optimum DC bias level where higher throughput can be achieved is 9~dB. So, we use 9~dB DC-bias for optic OFDM in our simulations.

\begin{figure}[!ht]
	\centering
	\includegraphics[width=35pc]{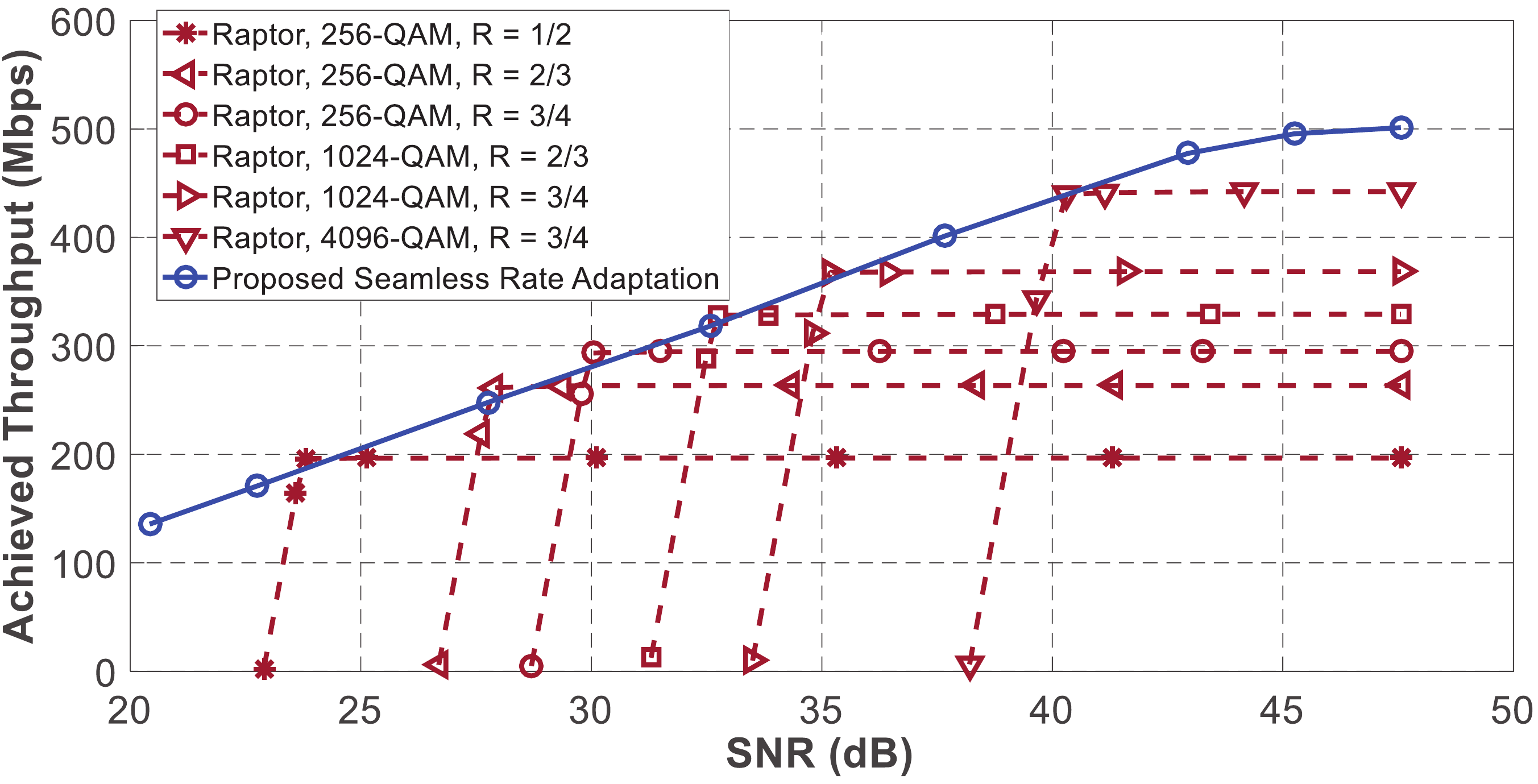}
	\caption{Achieved throughput obtained by utilizing proposed rate adaptive method and AMC method for SNR range (20.42 dB - 47.61 dB) inside the model room.}
	\label{fig_snr_throughput}
\end{figure}

\begin{figure}[!ht]
	\centering
	\includegraphics[width=35pc]{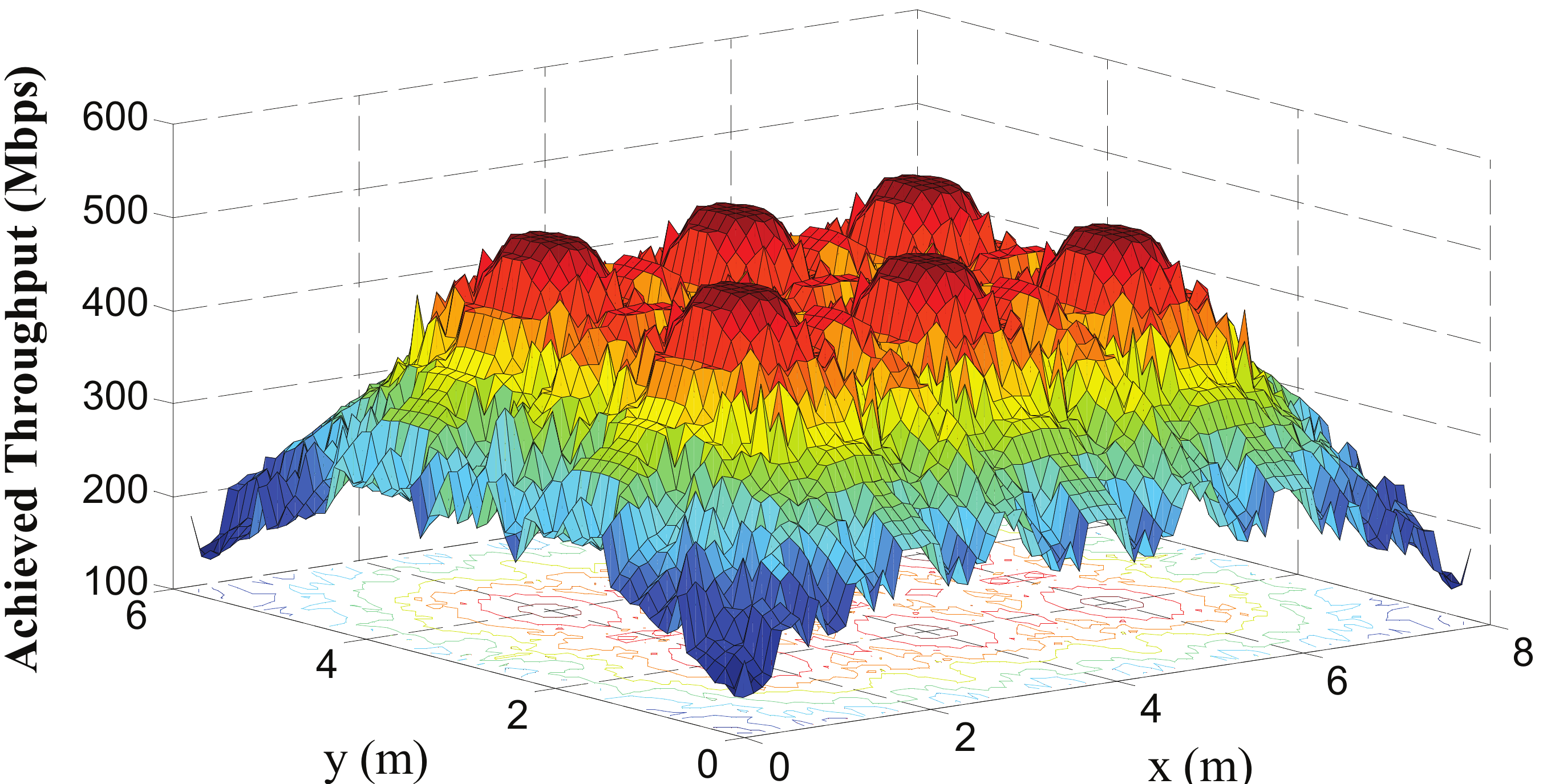}
	\caption{Distribution of achieved throughput obtained by proposed method inside the model room}
	\label{fig_throughput_distribution}
\end{figure}

Additionally, Fig.~\ref{fig_snr_throughput} illustrates the throughput performance comparison between proposed rate adaptive scheme with AMC method for 100~MHz bandwidth under highly variable SNR values which may occur because of mobility. For a fair comparison, considered modulation depths and code rates for AMC method in Fig.~\ref{fig_snr_throughput} are chosen to obtain the highest possible throughput values. 
As it can be seen in the figure, AMC achieve staircase throughputs by switching among fixed code rates and modulation levels pairs. For the most of SNR values occur in the model room, achieve throughput with proposed rate adaptive method is higher than throughput of AMC method.
These results also indicate that without rate adaptation, system would be designed for the lowest SNR value in model room which is 20.42~dB and throughput would be 136~Mbps. However, with rate adaptation, throughput performance can increase up to 500~Mbps. In Fig.~\ref{fig_throughput_distribution} we give achieved throughput variation vs location inside of the model room for proposed method. The throughput changes between 136~Mbps~-~501~Mbps. 

\begin{figure}[!ht]
	\centering
	\includegraphics[width=35pc]{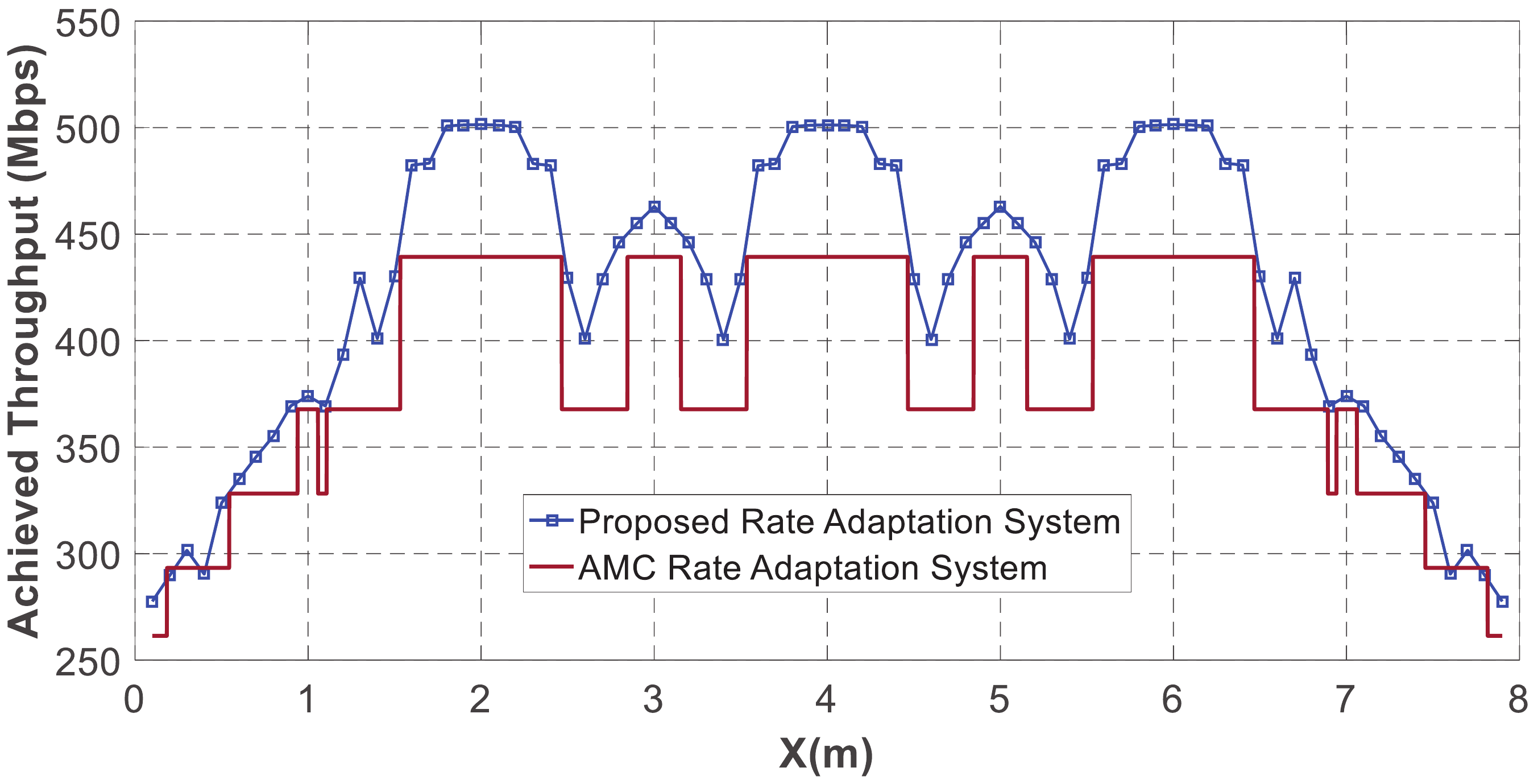}
	\caption{Achieved throughput for a scenario that a user walks along the x axis from (0.1m, 2.0m, 0.8m) to (7.9m, 2.0m, 0.8m) in considered model room}
	\label{fig_route_throughput_distribution}
\end{figure}

Finally, comparison between proposed and AMC methods in presence of mobility is done by a mobility scenario where receiver moves along \textbf{x} axis from (0.1m, 2.0m, 0.8m) to (7.9m, 2.0m, 0.8m) inside the model room. The results given with Fig.~\ref{fig_route_throughput_distribution} illustrate that proposed rate adaptation scheme offers about \%10.60 better throughput compared to AMC method. It should be stated that for all calculations for AMC method, it is assumed that receiver provides accurate CSIs to transmitter without any feedback delays. But proposed method does not need CSI at transmitter.

\section{Conclusion}
\label{Conclusion}
In this study, we propose a Raptor codes based rate adaptive method which adjusts transmission rate at receiver side and doesn't require CSI at transmitter side. 
Spectral efficiency of proposed method is analyzed and compared with AMC method by various simulation works. 
We consider a realistic communication system model for simulations by using commercially available LEDs and also fulfilling illumination requirements. 
To overcome multi-path effects, DCO-OFDM is used after determined DC bias level for higher throughput.
For a fair comparison, considered modulation depths and code rates for AMC method are chosen to obtain the highest possible throughput values.
Simulation results indicate that proposed rate adaptive system can achieve seamless rate adaptation and has higher spectral efficiency compared to AMC in presence of significant signal quality variations caused by mobility in model room.
Note that, given simulation results don't contain effect of CSI feedback transmission from receiver to transmitter side which is required for AMC resulting in higher spectrum deficiency.







\end{document}